\documentclass[11pt]{article}
\usepackage{amsmath}
\usepackage{amsfonts}
\usepackage{epsfig}
\newcommand{\ds}{\displaystyle}
\def\Vc{{\cal V}}

\usepackage{amsmath}
\usepackage{amsfonts}
\usepackage{epsfig}
\newcommand{\be}{\begin{equation}}
\newcommand{\ee}{\end{equation}}
\newcommand{\bea}{\begin{eqnarray}}
\newcommand{\eea}{\end{eqnarray}}

\DeclareMathSymbol{\mg}{\mathrel}{symbols}{"1D}

\newcommand{\ga}{\alpha}

\newcommand{\gx}{\xi}
\newcommand{\gm}{\mu}

\newcommand{\gl}{\lambda}
\newcommand{\gr}{\rho}

\newcommand{\gz}{\zeta}
\newcommand{\gp}{\pi}

\newcommand{\gF}{\Phi}

\newcommand{\cP}{{\cal P}}

\newcommand{\Li}{\mbox{Li}}

\newcommand{\ra}{\rightarrow}

\renewcommand{\Im}{\text{Im}\ }

\newcommand{\tabu}[2]{\begin{tabular}{#1} #2 \end{tabular}}

\newcounter{oldcounter}

\if@twoside\oddsidemargin25mm
\else
\begin{document} 
\begin{flushright}  
{DAMTP-2002-42}
\end{flushright}  
\vskip 2 cm 
\begin{center} 
{\Large {\bf Stability of the Higgs mass in theories with \\
\vspace{0.3cm}
extra dimensions$^\dagger$}} 
\\[0pt] 
\bigskip 
\vspace{0.93cm} 
{\large 
{\bf D.M. Ghilencea$^{a,}$\footnote{
{{ {\ {\ {\ E-mail: D.M.Ghilencea@damtp.cam.ac.uk}}}}}} and 
{\bf H.-P. Nilles$^{a,}$\footnote{
{{ {\ {\ {\ E-mail: Nilles@th.physik.uni-bonn.de}}}}}} }}}
\\ 
\vspace{0.9cm} 
$^a${\it DAMTP, CMS, University of Cambridge} \\
{\it Wilberforce Road, Cambridge, CB3 0WA, United Kingdom.}\\
\bigskip 
$^b${\it Physikalisches Institut der Universit\"at Bonn} \\
{\it Nussallee 12,  Bonn, 53115, Germany.}\\
\bigskip 
\end{center}
\vspace{2cm}
\begin{center}
{\bf Abstract}\\
\vspace{0.5cm}
\end{center}
We analyze the ultraviolet stability of the Higgs mass in 
recently proposed Kaluza-Klein models compactified 
on  $S_1/Z_2$ or  $S_1/(Z_2\times Z_2')$, both at the field theory
and string theory level. Fayet-Iliopoulos terms of $U(1)$ hypercharge
are shown to be of vital importance for this discussion. Models with 
a single Higgs doublet seem to be generically affected by quadratic 
divergences.
\bigskip 
\bigskip
\bigskip
\vfill{
{\noindent
$\dagger$ To be published in the Proceedings of Durham IPPP meeting, May 2001.
\newline 
Based on authors' talks at:\\
Durham IPPP meeting, May 2001, ``Phenomenology Beyond the Standard
Model'',  Durham, U.K.\\
Corfu Meeting, Sept. 2001, ``Across the Energy Frontier'', Corfu, Greece.}
}

\newpage
\noindent
\section{General remarks on Kaluza-Klein models}
\noindent
Perturbative calculations in quantum field theory
are often accompanied by the presence of divergences. 
In the absence of anomalies, these divergences do not lead to inconsistencies,
as their presence may be absorbed in the redefinition of the 
physical quantities of the model as for example  the coupling
constants of the theory.
There are cases such as the Higgs sector in the Standard Model 
where these divergences are quadratic in the scale $\Lambda$ or even quartic in 
the case of the vacuum energy. Therefore the physics of the Higgs sector is
strongly dependent on the ultraviolet details of the theory. 
Indeed, in the Standard Model (SM), at one-loop level:
\begin{equation}\label{sm}
m_H^2 \propto \alpha\, \Lambda^2
\end{equation}
While stabilising the value of the Higgs mass to small  values may be
done (in a one loop order), this procedure must be repeated to every
order in perturbation theory. This then brings a large amount of fine
tuning in the theory.

A standard procedure to avoid 
this problem of the Standard Model is to introduce
a supersymmetric extension of it, 
which is free of quadratic and quartic divergences, 
provided an additional condition (see later) 
for $U(1)_Y$ sector is respected.
However, since supersymmetry is not an exact symmetry in Nature,
a softly broken supersymmetry  \cite{Ferrara:1979wa}
at some scale $m_{susy}$ is actually good enough for 
the purpose of alleviating the ultraviolet behaviour
of the Higgs boson mass. 
In such a case the quadratic divergence in the Higgs sector
is alleviated  to one of (rather mild) logarithmic type.
In this case the following relation  emerges, 
with this solution 
to the UV sensitivity  of the Higgs mass to  hold beyond one loop order:
\begin{equation}\label{soft}
m_H^2 \propto   m^2_{Susy} \ln \frac{\Lambda}{m_{Susy}}
\end{equation} 
In the limit {$m_{Susy}\rightarrow \infty$} the quadratic divergence 
is however restored, hence the need for a low energy (softly)
broken supersymmetry. In this case the Higgs sector is only 
logarithmically sensitive to/dependent on the ultraviolet cut-off
of the theory. 

For supersymmetric model building purposes, 
condition  (\ref{soft}) is however not enough for the 
absence of ultraviolet divergences of the Higgs mass beyond that of 
logarithmic type in (\ref{soft}) or that
induced by the UV behaviour of the couplings of the theory.
The reason for this is that there exists a contribution from the 
Fayet Iliopoulos (FI) term  which brings in a 
quadratic dependence on $\Lambda$ of the Higgs mass. 
This contribution  is also proportional to the 
sum of the hypercharges of all massless complex scalars
\cite{Fischler:1981zk} of the model.
Therefore the sum of these hypercharges must be zero to ensure 
the absence of quadratic divergences, and this is
indeed the case of the Minimal Supersymmetric Standard Model (MSSM).

With  growing interest in the physics of large extra dimensions
various higher dimensional (supersymmetric) extensions of the 
standard model have been proposed. The simplest of these Kaluza-Klein models
involve one additional compact dimension with radius $R$ and employ
either an orbifold compactification $S_1/Z_2$ or $S_1/(Z_2\times Z_2')$
with mixed orbifold compactification and 
Scherk Schwarz breaking \cite{Scherk:1979zr}
of supersymmetry. One loop corrections to the Higgs mass have been
computed in these models in ref.\cite{Antoniadis:1999sd}, \cite{Delgado:1999qr},
\cite{Barbieri:2001vh}, \cite{Arkani-Hamed:2001mi}, 
\cite{Delgado:2001si}. Employing a specific regularisation scheme
one obtains a finite result:
\begin{equation}\label{higgsmassk}
m_H^2 \propto \frac{\alpha}{R^2} 
\end{equation}
with no explicit dependence on  $\Lambda$.
For particular models \cite{Barbieri:2001vh} it has been suggested that
the finite result for the Higgs mass holds to all orders in
perturbation theory \cite{Barbieri:2001dm}. 
This assertion is  not necessarily justified as we will see later on.

The absence of any explicit cut-off dependence in the one loop
expression (\ref{higgsmassk})
means that the models have  one parameter less and this 
will allow numerical predictions for the Higgs mass.
As with all effective models with additional (large) 
extra dimension(s), the 
new guise of the hierarchy problem is in this case that
of dynamically ``fixing'' the value of the compactification scale,
$1/R$  which in the aforementioned Kaluza-Klein models
is required/implicitly  assumed to be small, of TeV order.
Otherwise, for  $1/R$ large or of the order of the
cut-off, one recovers the quadratic divergence of the Standard Model
case, eq.(\ref{sm}), thus the importance of ``fixing'' $R$.
One can also say that 
the need for  small $1/R$ in Kaluza-Klein models 
is equally important to the need for a
low scale of supersymmetry breaking in supersymmetric models, 
$m_{susy}\approx 1 TeV$, see 
eqs.(\ref{soft}). This is because in both cases the Higgs (mass)$^2$ 
is proportional to the square of these quantities.
A very large compactification scale $1/R$ would 
require a fine tuning of  the couplings $\alpha$ to extremely small 
values,  such that the prediction of the Higgs mass remain within 
the electroweak scale region, according to eq.(\ref{higgsmassk}). 
This can then significantly affect the phenomenological viability
of the Kaluza-Klein  models considered.

For a given model one can define the  UV sensitivity of the Higgs mass  
(or amount of fine tuning to keep it close to electroweak
scale)   
\begin{equation}
\zeta_H \sim \frac{1}{m_H^2} \frac{d m_H^2}{d \ln\Lambda}
\end{equation} 
Other choices are possible, but the qualitative discussion below will
not change. The UV sensitivity for Kaluza-Klein models, if 
(\ref{higgsmassk}) holds beyond one-loop, is controlled  by
the UV behaviour of the couplings $\alpha$.  
The (gauge or Yukawa) couplings $\alpha$ of 
Kaluza-Klein models inevitably have some UV sensitivity  to 
the cut-off due to their additional dimension(s). 
A one-loop corrected coupling in (\ref{higgsmassk}) 
amounts to a two-loop order correction to the Higgs mass.
This one-loop  coupling is in general expected to be
linearly dependent  on the scale in 5D Kaluza-Klein
models. Therefore two loop  Higgs mass is expected to have 
linear UV sensitivity $\zeta_H\sim d \ln \alpha/d \ln \Lambda\sim
\alpha \Lambda R$,  even  though no apparent, explicit cut-off 
dependence is manifest at one-loop level, eq.(\ref{higgsmassk}). 
In a model by model approach it would be interesting to perform
a relative comparison of this 
{\it linear} sensitivity induced beyond one 
loop, but suppressed by an additional  power of the coupling to 
the logarithmic one  of  the MSSM case. 
This would be helpful in establishing accurately if an improvement from the 
MSSM case is  possible beyond one loop order.

Generic five dimensional N=1 supersymmetric 
Kaluza-Klein models constructed recently
fall in two classes: those whose low energy (massless) spectrum
is  that of the MSSM \cite{Delgado:1999qr}, \cite{Delgado:2001si} usually
compactified on $M^4\times S^1/Z_2$ and those 
compactified on $M^4\times S^1/(Z_2\times Z_2')$,
where this is just the SM spectrum \cite{Barbieri:2001vh}
Regardless of the  details of the phenomenological 
viability of these models,
one of their nice features is the presence of the electroweak
symmetry breaking triggered by the boundary conditions one
chooses for the 5D multiplets. Indeed the scalar potential in
these models can be  written in general as 
\begin{equation}\label{potential1}
\Vc(\phi)=\frac{1}{2} Tr \sum_{k=-\infty}^{\infty}\int \frac{d^4 p}{(2\pi)^4}
\ln\frac{p^2+m_{B_k}^2(\phi)}{p^2+m_{F_k}^2(\phi)}
\end{equation}
where the masses of the Kaluza-Klein states has the structure
\begin{equation}\label{mm}
m_{B_k}^2=\left[\frac{2}{{R}}\right]^2(k+q_{B})^2+M^2_{\phi}
\end{equation}
for bosonic states, with similar definition for the fermionic states.
For the case of \cite{Barbieri:2001vh}
$M_{\phi}=0$ with the field
dependence introduced via $q_{B(F)}$.  At field theory level the
apparent ``freedom'' of choosing the boundary conditions fixes the relative
values of $q_{B}$ and $q_F$. This in turns fixes the sign
of the scalar potential (\ref{potential1}) (and of its second
derivative) and thus the  presence  or absence of the 
electroweak symmetry breaking by one loop radiative effects. 
In the more general  context of string
theory, compactified on
$S_1/(Z_2\times Z_2)$ with Scherk Schwarz breaking of supersymmetry, 
it turns out \cite{Ghilencea:2001bv} that integer Kaluza-Klein  momenta 
correspond to bosons $q_B=0$ and  half-integer  momenta correspond
to fermions, $q_F=1/2$. This boundary condition would not lead to EW 
symmetry breaking by one loop effects.

The first class of models mentioned (see e.g. \cite{Delgado:2001si})
has some similarities with the  
MSSM, with the  nice benefit of a one-loop cut-off independent Higgs
mass, eq.(\ref{higgsmassk}) as compared to the MSSM case, eq.(\ref{soft}).
They also have electroweak symmetry breaking triggered by the effects of
Kaluza-Klein states. However, the couplings of the models run power-like
with the scale. One can adopt two views of this situation. One is that the
couplings (gauge or Yukawa) are just  an input/fixed from experiment, 
and thus no  UV sensitivity of $m_H$ exists in one loop order or
beyond, if the result (\ref{higgsmassk}) holds. The second view 
- and we share this point of view - is that in
going beyond one loop order, the couplings' ``running'' is
likely  to re-introduce  some UV sensitivity of the low energy 
physics\footnote{This second point of view would correspond to the 
Wilsonian interpretation of the low energy effective field theory.}
($m_H$), as we discuss in the following.
For a specific example \cite{Delgado:2001xr}
it was shown that Yukawa corrections to the Higgs mass induced by KK
states bring in for $m_H^2$ a  linear divergence at two loop level.  
This  linear divergence  can be re-absorbed in this case in  
the re-normalisation of Yukawa couplings, showing that  a 
non-renormalisation of the superpotential 
still holds at one loop order (despite breaking supersymmetry by a 
Scherk Schwarz mechanism). Due to the non-renormalisable character 
of the models, it is not  clear that all (beyond one loop) divergences 
may  systematically be  absorbed in the redefinition of physical 
quantities  of the theory. Even so,  
the conclusion is that   two-loop (Yukawa)
corrected  Higgs mass is a finite function of the  re-scaled,
linearly divergent, Yukawa  coupling. Then some UV sensitivity
of the Higgs mass will  still exist, brought in by the coupling' 
dependence on the UV physics, even though no additional explicit cut-off
dependence of $m_H$ is present. 

In  a sense the discussion above suggests  that Kaluza-Klein 
threshold corrections to the  Higgs mass (or scalar potential)
are related to the Kaluza-Klein threshold corrections
to the gauge/Yukawa couplings of the model. 
In some cases the UV (cut-off or scale) dependence of $m_H$ may 
then be redefined  in terms of that of the couplings of the theory. 
For this reason  it is sometimes said that,
if the couplings of the theory are kept {\it fixed}, just as a final  
input,  the Higgs mass is ``calculable''. This does not
mean  that $m_H$ would be UV insensitive. Indeed, if
(\ref{higgsmassk}) holds beyond one loop in a particular model, 
the UV sensitivity of $m_H$ in that order $(n)$ will still be (related
to) that of couplings of the model in the corresponding order (usually
$(n-1)$) of perturbation  theory.

The second class of models mentioned is that with $S_1/(Z_2\times Z_2')$
compactification. An explicit model is provided for example by 
the construction of ref.\cite{Barbieri:2001vh}
with the massless spectrum chosen to be that of the SM. The model  
has some supersymmetric features, since it 
is obtained by compactifying N=1 supersymmetric theory in 5  dimensions 
on $M^4\times S^1/(Z_2\times Z_2')$, with KK states to fall into
multiplets of this supersymmetry. However,   the massless
spectrum is that of the Standard Model with only one Higgs state. It
is thus not clear to what extent the initial supersymmetry is able to protect 
the model from divergences, to give a (one loop) finite Higgs mass of type
(\ref{higgsmassk}). In \cite{Ghilencea:2001ug} concerns have been
raised  on the  finiteness (beyond one loop) and on the
UV insensitivity or amount of 
fine-tuning of the Higgs mass in such models,  with further 
analysis  on the link with string theory in \cite{Ghilencea:2001bv}.
In \cite{Ghilencea:2001bw} it has been shown that the Higgs mass in
this model has a quadratic UV divergence already at one loop order and 
eq.(\ref{higgsmassk}) does not apply. The divergence of the Higgs mass
is a consequence of a divergent Fayet Iliopoulos (FI) term of
$U(1)$ hypercharge. Since it turns out that FI terms are very important
for the discussion of generic higher dimensional models
\cite{GrootNibbelink:2002wv}, let us briefly
describe in the next section the calculation of the FI term 
in  generic models with $M_4\times S_1/(Z_2\times Z_2')$ compactification.

\section{\bf The Fayet-Iliopoulos term.}
For the purpose of such evaluation \cite{Ghilencea:2001bw}
one can start with the 5D N=1
action describing a vector $V$ and a hypermultiplet $H$ \cite{Gunaydin}
and compactify it on $M_4\times S_1/(Z_2\times Z_2')$.
The  vector multiplet has components $V=(A_M,\lambda,\Phi,D^a)$ with
$A_M$ the vector field, $\lambda$ the Majorana gaugino, $\Phi$ a real
scalar field, and $D^a$ an $SU_R(2)$ iso-triplet of auxiliary scalars.
The hypermultiplet $H=(h_\alpha^i,\zeta_\alpha,F_\alpha^i)$ 
contains a bi-doublet of $SU(2)\times SU_R(2)$
$h_\alpha^i$ (scalars), a hyperino $\zeta_\alpha$ and auxiliary (bi-doublet)
scalars $F_a^i$. The parity assignments for the Kaluza-Klein states
are presented  in the Table  below
\begin{center}
\tabu{l | l | l | l | l | l | l | l | l | l }{
fields & 
$h^\ga_{n\,\ga}$ & $h^\ga_{n-\ga}$ & $\gz^\ga_n$ & $A_\gm^n$ & 
$A_5^n$ & $\gl_i^n$ & $\gF_n$ & $D^\parallel_n$ & 
$\vec{D}^\perp_n$ 
\\ \hline 
parities & $--$ & $++$ & $\ga-\!\ga$ & $++$ & $--$ & $i-\! i$ & 
$--$ & $++$ & $--$  
\\
modes $n$ & $ \geq 1$ & $\geq 0$ & $\geq 0$ & $\geq 0$ & $\geq 1$ & 
$\geq 0$ & $\geq 1$ & $\geq 0$ & $\geq 1$
}
\end{center}
with $\ga, i = \pm$. Of the three auxiliary scalars, that form a triplet $\vec D$ 
under $SU_R(2)$ of the $N = 1$ vector multiplet, 
two are odd under both parities 
$\vec D^\perp = (1 - \vec a_R \vec a_R^T) \vec D$, while the 
other one $D^\parallel = \vec a_R^T \vec D $ is even. 

The corresponding diagram of the FI contribution to the self energy 
of a scalar  is
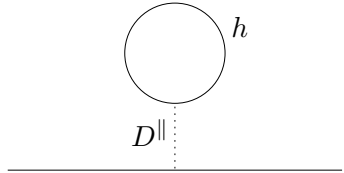
\begin{center}
\begin{picture}(0,0)%
\includegraphics{FI.pstex}%
\end{picture}%
\setlength{\unitlength}{2763sp}%
\begingroup\makeatletter\ifx\SetFigFont\undefined%
\gdef\SetFigFont#1#2#3#4#5{%
  \reset@font\fontsize{#1}{#2pt}%
  \fontfamily{#3}\fontseries{#4}\fontshape{#5}%
  \selectfont}%
\fi\endgroup%
\begin{picture}(3024,1519)(2089,-5773)
\put(4126,-4561){\makebox(0,0)[lb]{\smash{\SetFigFont{11}{13.2}{\familydefault}{\mddefault}{\updefault}
$h$
}}}
\put(3226,-5536){\makebox(0,0)[lb]{\smash{\SetFigFont{11}{13.2}{\familydefault}{\mddefault}{\updefault}
$D^\parallel$
}}}
\end{picture}

\end{center}
The dotted line corresponds to the auxiliary field $D^{||}$ of the Abelian
gauge multiplet in 4D. 

We investigate what happens to the FI term in the effective field
theory coming from 5D with a mass spectrum of complex scalars of the
hypermultiplet, in $S_1/(Z_2\times Z_2')$ compactification.
We take the charges of these scalars as $q_n^{++}=-q_n^{--}=1$
because they belong to complex conjugate representations.
The expression for the one loop contribution to the FI term 
whose diagram is presented above, is \cite{Ghilencea:2001bw}
\begin{equation}\label{ev}
\xi = \sum_{n, \ga} \, g\,  q^{\ga\ga}_n \, \int \frac{d^4 p_4}{(2\gp)^4} 
\frac{1}{p_4^2 + ({m_n^{\ga\ga}})^2 + m^2}
\end{equation}
where $m_n^{\ga\ga} = 2n/R$ and the sum for $\ga = +$ is over 
$n \geq 0$, while for $\ga = -$ over $n > 0$ in agreement with parity
assignments of the above table.

The expression in (\ref{ev}) is considerably more complicated than 
the usual 4D case,  each scalar mode of the infinite sum brings in a 
quadratic divergence, and a careful evaluation is required.
To compute (\ref{ev}) one can transform the sum into an integral in
the complex plane \cite{GrootNibbelink:2001bx} as a regularisation
procedure. With this regularisation  we find an 
integral over the compact dimension of the following
 structure \cite{Ghilencea:2001bw}
\begin{equation}\label{xi}
\gx^{} = g\,  \int \frac{d^{D_4} p_4}{(2\gp)^{D_4}} 
\int_\ominus \frac{d^{D_5} p_5}{2 \gp i}
\left[
\frac{\cP^{++}(p_5)}{p_4^2 + p_5^2 + m^2} - 
\frac{\cP^{--}(p_5)}{p_4^2 + p_5^2 + m^2}
\right].
\end{equation}
where we have introduced the ``pole functions'' \cite{GrootNibbelink:2001bx}
$\cP^{++}$ and $\cP^{--}$.
For $\Im p_5 > 0$, the pole functions may be written as \cite{Ghilencea:2001bw}
\begin{equation}
\cP^{\ga\ga} = \frac{1}{2} \left[
- \frac i2 \gp R + \frac \ga{p_5} 
- \gr_-(p_5) \right],\quad
\quad
\gr_\ga(p_5) = i\gp R \frac {e^{i\gp R\, p_5}}{1 + \ga e^{i\gp R \, p_5}}.
\label{polefunS22>}
\end{equation}
For the case  $\Im p_5 < 0$  one finds \cite{Ghilencea:2001bw}
\begin{equation}
\cP^{\ga\ga} = \frac{1}{2} \left[ \frac i2 \gp R + \frac \ga{p_5} + \gr_-(-p_5)
\right],\quad
\label{polefunS22<}
\end{equation}
When  integrated, the first term in $\cP^{\alpha\alpha}$ 
accounts for genuine 5D divergences, 
the second $(\alpha/p_5)$ for 4D divergences while the remaining part
gives the finite contributions.

Making use of  the expressions of $\cP^{\alpha\alpha}$ in the expression
(\ref{xi}) we find that the result for $\xi$  is:
\begin{equation}
\xi = g\, \frac {1}{2\gp i}\,  \int \frac{d^{D_4} p_4}{(2\gp)^{D_4}} 
\int_\ominus d^{D_5} p_5\, 
\frac{1}{p_5} 
\frac {1}{p_4^2 + p_5^2 + m^2} 
= g\, \int \frac{d^{D_4} p_4}{(2\gp)^{D_4}} \frac {1}{p_4^2 + m^2}.  
\label{FItermzero}
\end{equation}
We can safely take $D_5=1$ and remove the contour integration in
(\ref{FItermzero}) to obtain an expression similar to that in  the 
4D case.
This just shows that the FI term at one loop  is simply given by the 
contribution of the  massless complex scalar ($m \ra 0$) and leads to
a quadratic divergence of the zero mode scalar mass (Higgs). 
 
Further gauge corrections exist to the self energy of the zero mode scalar
(identified as a Higgs state with continuous line in the
diagrams below). They can be computed using the same 
procedure as for the FI term. The
corresponding Feynman diagrams are:
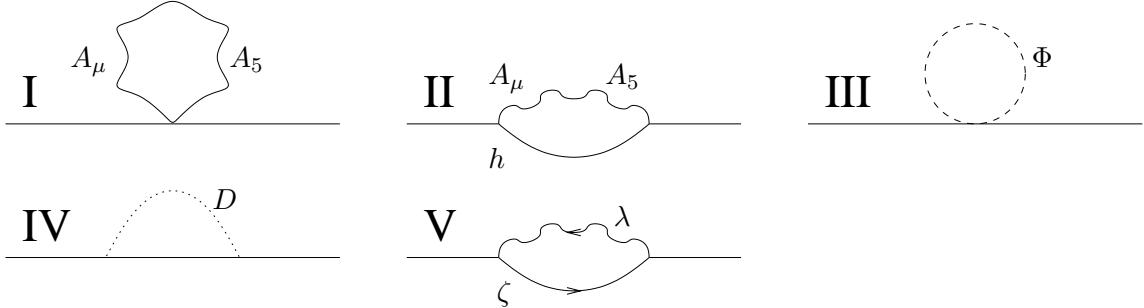
\begin{center}
\begin{picture}(0,0)%
\includegraphics{GaugeCon.pstex}%
\end{picture}%
\setlength{\unitlength}{2763sp}%
\begingroup\makeatletter\ifx\SetFigFont\undefined%
\gdef\SetFigFont#1#2#3#4#5{%
  \reset@font\fontsize{#1}{#2pt}%
  \fontfamily{#3}\fontseries{#4}\fontshape{#5}%
  \selectfont}%
\fi\endgroup%
\begin{picture}(10224,2754)(2089,-7403)
\put(7501,-5386){\makebox(0,0)[lb]{\smash{\SetFigFont{11}{13.2}{\familydefault}{\mddefault}{\updefault}
$A_5$
}}}
\put(7576,-6661){\makebox(0,0)[lb]{\smash{\SetFigFont{11}{13.2}{\familydefault}{\mddefault}{\updefault}
$\gl$
}}}
\put(3976,-6511){\makebox(0,0)[lb]{\smash{\SetFigFont{11}{13.2}{\familydefault}{\mddefault}{\updefault}
$D$
}}}
\put(4126,-5236){\makebox(0,0)[lb]{\smash{\SetFigFont{11}{13.2}{\familydefault}{\mddefault}{\updefault}
$A_5$
}}}
\put(2701,-5236){\makebox(0,0)[lb]{\smash{\SetFigFont{11}{13.2}{\familydefault}{\mddefault}{\updefault}
$A_\gm$
}}}
\put(6526,-7336){\makebox(0,0)[lb]{\smash{\SetFigFont{11}{13.2}{\familydefault}{\mddefault}{\updefault}
$\gz$
}}}
\put(6451,-6136){\makebox(0,0)[lb]{\smash{\SetFigFont{11}{13.2}{\familydefault}{\mddefault}{\updefault}
$h$
}}}
\put(11326,-5236){\makebox(0,0)[lb]{\smash{\SetFigFont{11}{13.2}{\familydefault}{\mddefault}{\updefault}
$\gF$
}}}
\put(6451,-5386){\makebox(0,0)[lb]{\smash{\SetFigFont{11}{13.2}{\familydefault}{\mddefault}{\updefault}
$A_\gm$
}}}
\end{picture}

\end{center}
A wavy line stands for  a gauge field ($A_\gm, A_5$), 
a wavy line with an arrow stands for gaugino $\gl$, 
a line with an arrow -  a hyperino $\gz$, 
a dashed line a real scalar $\gF$ and a dotted line is an 
auxiliary field $D^a$. On the orbifold $S^1/(Z_2 \times Z_2')$ 
these fields are classified as even or odd under both parities.

The result of evaluating the above contributions to the zero mode
scalar   is
\begin{equation}
-i \Sigma_G=i \frac{7 g^2}{16 \pi^4} \left[\frac{1}{R/2}\right]^2\zeta(3)
\end{equation}
Since there is no quadratic divergence in this result, one can safely conclude
that it is not possible to cancel the quadratic divergence of the
Fayet Iliopoulos term to the scalar mass.  

The quadratic divergence of the  correction to the Higgs massless
mode eq.(\ref{FItermzero}) holds for any finite $R$, since it is independent 
of the radius $R$ of the compact dimension. Therefore, we conclude 
that it is also true in the limit $R \ra \infty$. 
This signals that the boundary condition of the orbifolding 
procedure is not removed in limit $R \ra \infty$. The 
de-compactification limit is thus not smoothly connected
to a supersymmetric theory (obtained through compactification
on a circle).

It is perhaps worth mentioning here that 
finite temperature arguments  are sometimes used  
to justify a soft ultraviolet behaviour of the  Higgs mass in 4D effective models
with one additional compact dimension (5D N=1 supersymmetric
model to start with). They proceed as follows. 
The inverse radius of the compact dimension  corresponds to the 
temperature. Supersymmetry in the initial 5D model 
then ensures a vanishing of the potential/Higgs mass in 
the $R \ra \infty$ (or $T=0$) limit. 
Finite values of $R$ would then trigger only finite corrections 
to the scalar potential/Higgs mass, by analogy with finite temperature calculations
of the free energy, since in going from $T=0$ to $T\not=0$
only finite corrections could appear, if the limit 
$T \ra 0$ ($R \ra \infty$) is smooth. However it seems to be impossible
to reconcile this mechanism 
with the phenomenological requirement
to obtain a chiral spectrum in 4D, while smoothly varying the radius
from $R\ra \infty$ of the N=1 supersymmetric limit in 5D
to a broken supersymmetry phase.  An orbifold-like compactification is 
necessary, but then the limit $R\ra\infty$ on the orbifold 
does not restore the full initial supersymmetry at the fixed points,
as we just discussed in the example  above.
In other words finite temperature arguments may apply to a
circle compactification of the extra dimension (non-chiral models), 
but not necessarily to a manifold with fixed points 
(orbifold compactifications).

At an early stage of the discussion, a further explanation for the 
finiteness of the Higgs mass was the argument that a counterterm for
the Higgs mass was forbidden by the symmetries of the theory.
Various symmetries have been suggested in that direction. 
The result of ref. \cite{Ghilencea:2001bw}, however, suggests a new 
counterterm in the theory that is closely connected to the Higgs mass.
The analysis of (localized) anomalies at the fixed points
\cite{Scrucca:2001eb}, \cite{Arkani-Hamed:2001is},
\cite{Pilo:2002hu}, \cite{Barbieri:2002ic}
clarifies the nature of exact symmetries of the model, which in fact
allow a counterterm for the Higgs mass in the
$S_1/(Z_2\times Z_2')$ model \cite{Barbieri:2002ic}.

The conclusion we draw from the above calculation is that 
a generic  5D N=1 supersymmetric model,  with a compactification 
$S_1/(Z_2\times Z_2')$ with massless sector identical to that of the  
SM will have a quadratic divergence for the Higgs mass at one-loop
level\footnote{This seems in disagreement with findings in \cite{Pilo}.}.  
In such case \cite{Barbieri:2001vh}  an improvement 
from the SM UV  behaviour of eq.(\ref{sm}), is not possible.
A suggestion  to avoid such problem may be to introduce an  
additional Higgs state as zero mode,  at the price of bringing an extra
parameter with implications on the predictive power of the model.

\section{Kaluza-Klein models - the view from string theory}
What is the situation in string theory? Afterall this is supposed to
provide a more consistent framework for investigating corrections to
the Higgs mass and scalar potential. At string level, the latter has a
correspondent in the vacuum energy, which receives
corrections from the string Kaluza-Klein and winding modes as well as
massive states.

Comparing the scalar potential of Kaluza-Klein models with the results from string
theory is not straightforward. The models must have  similar spectrum, orbifold
compactification and supersymmetry breaking mechanism. 
The Kaluza-Klein models must also be free of anomalies not only in 
4 dimensions, but also in their 5D embedding.
This is important for it may happen that anomalies in 5D, localised at
the  fixed points be present in some Kaluza-Klein models 
considered. In string theory their absence is ensured 
by some variation of  Green-Schwarz mechanism or by the twisted 
states, whose existence/absence in phenomenological studies of 
Kaluza-Klein models has so far been somewhat arbitrary.
It has been shown \cite{Scrucca:2001eb}, (see also 
\cite{Arkani-Hamed:2001is}) in a field theory approach that this is a 
non-trivial issue and that for some models anomalies   localised at
the fixed points may be present in 5D, with implications for the 
overall consistency of the models. We thus assume that these 
difficulties are overcome, and one is able to identify a consistent 
Kaluza-Klein model which is indeed the low energy limit of a heterotic 
or type I string model.

In the string context the phenomenological requirement that $R$ be 
of particular size requires a dynamical mechanism to fix the point 
in  moduli space giving this value. Further, the couplings receive 
some string scale  dependence/sensitivity, thus the
need for including the (string) threshold corrections to the 
gauge or Yukawa couplings for the model considered. 
Such threshold corrections are due to Kaluza-Klein and winding states
associated with the extra compact dimension(s) considered.  
Our previous discussion at the level of Kaluza-Klein models on the link
between UV sensitivity of the couplings and of the Higgs mass/scalar
potential has then a correspondent  at the string level. The  (string) 
threshold corrections  to the gauge/Yukawa couplings may be related  
to those to the vacuum  energy \cite{Ghilencea:2001bv}.
Therefore the UV sensitivity of the Higgs mass/vacuum energy and of the 
gauge couplings are related at string level as well.

Since most of the results of 5D N=1 supersymmetric Kaluza-Klein models 
of growing phenomenological interest  invoked a Scherk Schwarz mechanism 
for supersymmetry breaking, we consider such a case at the 
 level of (heterotic) string theory. The case of type I string is only 
briefly addressed.   We present  the results \cite{Ghilencea:2001bv}
for the vacuum energy/cosmological constant  for a $S_1/(Z_2\times Z_2')$ orbifold 
compactification,  although the results are rather generic to other 
compactifications. We limit ourselves to listing some comparative
features of the string results, to stress the ``regularisation'' role
of the winding modes and modular invariance.

Consider therefore a heterotic orbifold model with N=1
supersymmetry in four dimensions, which has untwisted N=4 and 
N=2 sub-sectors. Supersymmetry is broken in these sub-sectors of the orbifold
by a Scherk-Schwarz mechanism, by the following choice of the shifts of the
internal charges \cite{Antoniadis:1990ew}: 
\begin{equation}\label{ssmech}
n\ra n, \qquad m\ra m+p-\frac{1}{2}\, n, \qquad
p\ra p-n
\end{equation}
where $n,m,p$ are respectively the winding and  momentum numbers
with respect to the compactified direction, while p is the internal
$U(1)$  fermion charge.  
A Scherk-Schwarz mechanism with respect to the untwisted planes
of the orbifold will break N=2 and N=4 supersymmetry 
of the completely untwisted sector to N=0, while the N=1
sub-sector is left untouched (no mass shifts in this sub-sector).

Detailed calculations of the vacuum energy in string models with 
such a mechanism for supersymmetry breaking give the result of
eq.(\ref{sf}) below, which unlike that of Kaluza-Klein models, 
is well defined (finite) in the ultraviolet region. The only  model 
dependence of the result is manifest  in the coefficients 
$\gamma_N$. It is important to remark at this stage that the 
limit of integration in eq.(\ref{sf}) from deep ultraviolet $t \geq 0$ 
is introduced by  the presence of winding states. Essentially this comes about in
the following way. Initially the vacuum energy is defined as equal to
two sums over winding and momentum modes of an integral over the fundamental
domain   ${\cal F}_2$  of the string.
A technical argument \cite{O'Brien:1987pn}
enables one to re-define the summation indices into linear 
combinations of winding and momentum modes. One of these re-defined
sums when applied to  the fundamental domain of integration ``unfolds'' it to the
half  strip integration region defined 
by\footnote{Further, the integral over $\tau_1$  essentially projects all massive 
states to keep only those of equal mass levels (physical)
and one is left with the integral over $\tau_2$ only, eq.(\ref{sf}).} 
${\cal H}=\{\tau|-1/2\leq \tau_1 \leq 1/2; 0 \leq \tau_2 < \infty\}$.
In this ``unfolding'' procedure the winding modes' contribution is
essential, and they ensure that eq.(\ref{sf}) is well defined 
in the deep ultraviolet.
The second (remaining) sum is over  the integer $p$
which is a mixture of (Poisson re-summed) Kaluza-Klein and  
of winding numbers and it is  present in the formula below: 
\def\Li{{\cal L}i}
\begin{equation}\label{sf}
V_h(R)= R \int_{0}^{\infty} \frac{d\,\tau_2}{\tau_2^{7/2}} 
\sum_{p>0}^{}\sum_{N\geq 0} \left\{1-(-1)^p\right\}
e^{-\frac{\pi (R/2)^2}{
\tau_2} p^2} e^{-4 \pi \tau_2 N } \gamma_N
\end{equation}
This gives the full (heterotic) string result for the vacuum energy,
generic to orbifold compactifications  
with Scherk Schwarz breaking of supersymmetry in the untwisted sector.
The result is finite and no regularisation is required 
for the UV region $\tau_2\ra 0$. In this region 
most  contributions are exponentially suppressed in (\ref{sf}), 
except those of small radius (string units)  which
may still contribute. The latter may be interpreted as 
corresponding to Kaluza-Klein states of mass larger than the string scale, 
$k/R > M_{string}$. The presence  of such states in the string framework is 
justified, unlike the case of the (effective) field theory
where they may also be manifest, with a less clear physical  meaning
\cite{Ghilencea:2001ug}.

After some algebra the  (heterotic) string result can be
written as (using $x_N=exp(-4\pi R N^{1/2})$ 
and ${\cal L}i_n(x)=\sum_{k\geq 1} {x^k}/{k^n}$)
\begin{eqnarray}
V_h(R)&=&\ds{\frac{93\zeta(5)\gamma_0}{64\pi^2R^4}+
\frac{4}{R^2}\sum_{N> 0}\ \gamma_N\ N \left[\Li_3(x_{N})+
\frac{3}{4\pi R \sqrt N}\ \Li_4(x_N)+\frac{3}{16\pi^2 R^2 N}\ 
\Li_5(x_N)\right.}\nonumber \\
&&\ds{\left.\ -\Li_3(-x_N)-\frac{3}
{4\pi R \sqrt N}\ \Li_4(-x_N)-
\frac{3}{16\pi^2 R^2 N}\ \Li_5(-x_N)\right]\ ,}\label{polylog}
\end{eqnarray}
with the leading term proportional to $\gamma_0/R^4$ and remaining
terms  exponentially suppressed. $\gamma_0$  vanishes if the
ground state is supersymmetric. The model dependent coefficients
$\gamma_N$ have contributions from   $N\!=\!4$ and $N\!=\!2$
sub-sectors  of the initial orbifold, broken to $N=0$ by Scherk 
Schwarz mechanism.

An intriguing feature of the heterotic result eq.(\ref{sf}) is that  
it is close to the  type I result for vacuum energy, with Scherk 
Schwarz supersymmetry breaking.
The type I  result is \cite{Ghilencea:2001bv} (for full references and
review see \cite{typeI})
\begin{equation}\label{typeI}
V_{I}(R)=\int_{0}^{\infty}\frac{dt}{t^3} \sum_{n\in Z} (-1)^n
e^{-\pi t n^2/R^2} \sum_{N\geq 0}
\tilde\gamma_N e^{-\pi t N}\bigg\vert_{reg.}
\end{equation}
Unlike  the heterotic string case, a 
regularisation  is necessary in type I string case.
In (\ref{typeI})  ``reg'' stands for a regularisation of the
divergence introduced by the presence of  $n=0$ (for $N=0$) 
Kaluza-Klein state.  The  need for a regularisation  in type I 
case is not  new, it is manifest in other
calculations as well (for example threshold corrections  
to the gauge couplings 
\cite{Antoniadis:1999ge}). This is due  to  the
absence of modular invariance in these models which at heterotic level
ensures a finite result.  The absence of winding
modes requires a regularisation procedure,
and a  similarity of type I results to those in Kaluza-Klein models
with momentum modes only (no windings), will also 
exist. It is considered  that the  state $n=0$ is not
present if eq.(\ref{typeI}) is regularised in the transverse closed
string channel and one-particle irreducible diagrams are subtracted
\cite{Bachas:1996zt}.
Additional arguments for  such a  regularisation  and the absence of
the divergence when summing over all open string diagrams, are
represented  by the (requirement of)  tadpoles cancellation 
\cite{Antoniadis:1999ge}.

For  comparison of the (heterotic) string result with that of 
Kaluza-Klein models, we note that the former, eq.(\ref{sf}) can be re-written
after introducing a $p=0$ state and then Poisson re-summing over $p$
\begin{equation}\label{er}
V_h(R)=\int_{\epsilon^2 \alpha'}^{\infty} \frac{d \,z}{z^{3}}
\sum_{s=-\infty}^{\infty}
\left\{ e^{-\frac{\pi z}{(R/2)^2} s^2} -e^{-\frac{\pi z}{(R/2)^2} 
(s+1/2)^2}\right\} \sum_{N\geq 0} e^{-\pi z N} \gamma_N\ .
\end{equation} 
This expression is very similar  to that of {\it regularised} 
Kaluza-Klein models (except the summation over massive modes $N$) 
and that of type I case, eq.(\ref{typeI}). 
The presence of the lower limit of integration $\epsilon^2\alpha'$ in 
(\ref{er})  is needed for  bosonic and fermionic terms under
the integral be each well defined in the UV (for $N=0$, $z\ra 0$)
\cite{Ghilencea:2001bv}.   Therefore a regularisation is required
for a field theory interpretation/limit.
Strictly speaking, the summation index $s$ in (\ref{er}) is a
``mixture'' of Kaluza-Klein  number and (Poisson re-summed) 
winding number, which gives for the contribution to $V$ of the
momentum modes alone a further constraint $\vert s\vert \leq
(M_S R)/\epsilon$. This arises because the cut-off in (\ref{er})
$\epsilon^2\alpha'$ excludes the deep ultraviolet ($\tau_2\ra 0$)
 momentum region, to  retain under the integral over $z$ only the
 momentum modes with mass range $1/z<1/(\epsilon^2 \alpha')$.
In the field theory case with an infinite string scale ($\alpha'=0$), 
this constraint is removed. One assumes that {\it infinitely} many winding modes (and 
their effects) are entirely decoupled. Then  from
(\ref{ssmech}) at zero windings
$n=0$ a ``discrete shift symmetry'' emerges for Kaluza-Klein level $k\ra k+p$,
to require one sum over the whole Kaluza-Klein tower. 
A result with structure   close to that of the full  heterotic string 
result (\ref{sf}), (see also (\ref{polylog}), (\ref{er}))  - which includes winding
 modes' effects  - is then  obtained. In such case,
the field theory results and heterotic/type I  string results are similar
if:  in  eqs.(\ref{polylog}), (\ref{er})   one makes 
the formal replacement $N\rightarrow M_\phi^2$   and ignores the sum 
over the massive modes $N$. In this case the mass of KK states of 
field theory eq.(\ref{mm}) will appear  in the exponents in 
the integral (\ref{er}) and in (\ref{polylog}).

Since a field theory regime corresponds to a string calculation
with an infinite string scale, the string and field theory results 
may in general  be matched exactly only in this limit. 
Additional finite string effects which 
vanish in this limit  \cite{dmgsgn} may not be recovered by 
the field theory calculation.
In general, in the absence of a full string calculation to compare with,
one cannot state, on  field theory grounds,  that the {\it full} UV
behaviour of  a model with extra-dimensions is that found by using momentum modes
only.  Winding modes, although massive in the
field theory limit $\alpha'\ra 0$,  can  have additional
effects. Their presence enables the symmetries of the string 
(modular invariance) which require integrations take place 
over the fundamental domain, ensuring an UV finite (heterotic 
string) result. They then play a (regularisation) role in  the UV behaviour 
of the string  models in the limit $\alpha'\ra 0$. In some cases, for  
gauge couplings  \cite{dmgsgn}, they also control the exact value of  
the coefficient of the divergence the string result has 
in $\alpha'\ra 0$ limit, which is just the limit  of the field theory regime.
To conclude, to find the {\it full} UV behaviour of a model we 
argued that from a string
theory perspective it is more appropriate to consider 
the limit  $\alpha'\ra 0$ on the 
full string result. This should unambiguously determine the 
UV limit of the model, without being subject to a regularisation 
scheme dependence induced by field theory calculations.

\section{Conclusions}
We  have seen that the request for a better UV  behaviour 
(than in the SM, eq.(\ref{sm})) of the Higgs mass,  in models with
extra dimensions, leads to serious constraints on the spectrum and
the interactions of the theory. This should not be too surprising
as the higher dimensional theory is non-renormalisable.
We discussed the ultraviolet sensitivity of Higgs mass in Kaluza-Klein
models compactified on $S_1/Z_2$ and $S_1/(Z_2\times Z_2')$. 
For the latter class of models with massless modes as in the SM with
one Higgs state, a FI term is generated, which introduces a quadratic 
divergence for the mass of the Higgs particle.
In the more general  context of a string theory calculation 
for the vacuum energy,  the ``regularisation role'' of the winding 
modes was emphasized, as this ensures the finiteness of the 
(heterotic string) result.  This is unlike the  type I string case  
and Kaluza-Klein effective models, which do not have winding
contributions to the vacuum energy/scalar potential 
and where a regularisation is required.
To find the right UV behaviour of a model we argued it is more
appropriate to consider the limit $\alpha'\ra 0$ of the full string
result, which  should unambiguously determine the full  
UV behaviour of the model.

The requirement of mild UV sensitivity of the Higgs mass leads to 
significant constraints on the possible higher dimensional extensions
of the $SU(3)\times SU(2)\times U(1)$ standard model. FI terms of
U(1) hypercharge seem to play a central role in this discussion,
most notably in the (non-supersymmetric) $S_1/(Z_2\times Z_2')$
case. Even in the $N=1$ supersymmetric models, as e.g.
$S_1/Z_2$, divergent FI-terms localized at the orbifold fixed points
\cite{GrootNibbelink:2002wv} could lead to a destabilization of
the theory. Generically, these problems can be solved by a sufficient 
amount of supersymmetry and the presence of a second
Higgs multiplet with opposite hypercharge. This then leads us back to 
models very similar to the MSSM with softly broken supersymmetry.\\

\noindent
{\bf Acknowledgements}. 
We would like to thank S. Stieberger and S. Groot Nibbelink for
their  collaboration.  Work supported in part by a UK PPARC SPG 
research grant and 
the  European Community's Human Potential
Programme under contracts HPRN-CT-2000-00131 Quantum Space-time,
HPRN-CT-2000-00148 Physics Across the Present Energy Frontier and
HPRN-CT-2000-00152 Supersymmetry and the Early Universe.

\small{
}
\end{document}